# Evidence for dominant Pauli paramagnetic effect in the upper critical fields of a FeTe$_{0.6}$Se$_{0.4}$ superconductor

Seunghyun Khim[1], Jae Wook Kim[1], Eun Sang Choi[2], Yunkyu Bang[3], Minoru Nohara[4,5], Hidenori Takagi[5,6,7] and Kee Hoon Kim[1,6*]

[1] FPRD & Department of Physics and Astronomy, Seoul National University, Seoul 151-742, Republic of Korea
[2] NHMFL, Florida State University, Tallahassee, FL 32310, USA
[3] Department of Physics, Chonnam National University, Kwangju 500-757, Republic of Korea
[4] Department of Physics, Okayama University, Okayama 700-8530, Japan
[5] JST, TRIP, 5 Sanbancho, Chiyoda, Tokyo 102-0075
[6] Department of Advanced Materials, University of Tokyo, Kashiwa, Chiba 277-8561, Japan
[7] RIKEN, Advanced Science Institute, Wako, Saitama 351-0198, Japan

Abstract

We investigate temperature-dependence of the upper critical fields $H_{c2}(T)$ of a superconducting FeTe$_{0.6}$Se$_{0.4}$ single crystal by measuring resistivity in static magnetic fields up to 45 T. The observations of strong bending in the $H_{c2}^{ab}(T)$ curves and nearly isotropic $H_{c2}^{ab}(0) \approx H_{c2}^{c}(0) \approx 48$ T support the presence of strong Pauli paramagnetic effect. We show that the Werthamer-Helfand-Hohenberg formula considering the Pauli limiting and the spin-orbit scattering together can effectively describe both $H_{c2}^{ab}(T)$ and $H_{c2}^{c}(T)$ curves. The enhancement in quasi-particle density of states or increased scattering resulting from Te(Se) vacancy or excess Fe is discussed as a possible origin for the manifesting Pauli paramagnetic effect.

# I. INTRODUCTION

The observation of high temperature superconductivity in iron-pnictides has triggered a surge of research activity in recent years to investigate their basic superconducting properties and pairing mechanism.[1] The upper critical field $H_{c2}$ is one of fundamental superconducting parameters that provide clues to the pairing mechanism as well as the pairing strength. Moreover, the temperature-dependence of $H_{c2}$, $H_{c2}(T)$, and its anisotropy reflect the underlying electronic structure and delivers the valuable information on a microscopic origin for pair-breaking, which can be in turn important for application purposes.

In this respect, $H_{c2}(T)$ has been extensively studied in various forms of iron-pnictides ranging from the '1111' represented as $RE$FeAsO ($RE$ = rare earth) to the '122' system like $A$Fe$_2$As$_2$ ($A$ = alkali metal). Possibly due to the required, large field scale and scarcity of single crystals, the investigations on the '1111' system are still limited but have shown the existence of anisotropy between $H_{c2}$ in an $ab$-planar field $H_{c2}^{ab}$ and in a $c$-axis field $H_{c2}^{c}$ near superconducting transition temperature $T_c$.[2-6] Moreover, $H_{c2}^{ab}$ and $H_{c2}^{c}$ increase almost linearly or sublinearly with decreasing temperature near $T_c$, resulting the maximum slope change, $- dH_{c2}^{ab}/dT_c|_{max} \sim 9$ - 11 T/K. These characteristics of $H_{c2}$ curves support the multiband effect in the system. The '122' system also shows quite linear or sublinear increase of $H_{c2}^{ab}$ and $H_{c2}^{c}$ as well as their anisotropy near $T_c$, consistent with the multiband scheme. The maximum slope change $- dH_{c2}^{ab}/dT_c|_{max}$ in the '122' system is much smaller than the '1111' system, showing $\sim 3$ - 6 T/K.[7-10] In the orbital-limiting scenario, the expected $H_{c2}^{ab}(0)$ thus can be as high as 150~300 T in the '1111' system while it is about $\sim 80$ - 120 T in the '122' system.

In reality, however, most of the existing data for the $H_{c2}^{ab}(0)$ in the '122' system are smaller than 60 T. This experimental situation is also related with the fact that anisotropy ratio between $H_{c2}^{ab}$ and $H_{c2}^{c}$ decreases with decreasing temperature in most of the iron pnictides.[7-10] Thus, most of '122' single crystals and thin films covering both hole-doped (Ba,K)Fe$_2$As$_2$ and electron-doped Sr(Fe,Co)$_2$As$_2$ have shown nearly isotropic $H_{c2}(0)$ behavior. The existence of such isotropic $H_{c2}(0)$ in the '122' materials with the cylindrical Fermi surface is quite unusual and is in sharp contrast to the case of layered cuprates. Although the band warping in the cylindrical surface or multiband effects has been discussed as a possible origin, those mechanisms alone might not be enough to explain the existence of isotropic $H_{c2}(0)$ insensitive to doping level and degree of disorder.

The iron-chalcogenides Fe(Te,Se) with the PbO-type structure is yet another new Fe-based superconductors with $T_c$ = 8.0 - 14.5 K.[11-14] Its structure is characterized with the simple planar sheets of tetrahedrally coordinated Fe, which is common to the iron-pnictides superconductors. The Fermi-surface is composed of cylindrical hole and electron pockets, similar to those of the iron-pnictides.[15-16] In view of the similarity in the electronic structure, the study of $H_{c2}(T)$ and its anisotropy in the Fe(Te,Se) is expected to provide useful comparison to the '122' system and elucidate the origin of the isotropic $H_{c2}(0)$. The first attempt to investigate the $H_{c2}(T)$ in polycrystalline FeSe$_{0.25}$Te$_{0.75}$ found a strong bending of the $H_{c2}$ curve with lowering temperature, indicating the Pauli paramagnetic effect.[17] A subsequent measurement on a single crystal Fe$_{1.11}$Te$_{0.6}$Se$_{0.4}$ showed a weak anisotropy in the $H_{c2}(0)$, which was interpreted as a possible band warping effect, similar to the '122' case.[18]

In this study, we have used static magnetic fields up to 45 T to determine the resistive $H_{c2}^{ab}(T)$ and $H_{c2}^{c}(T)$ curves of a FeTe$_{0.6}$Se$_{0.4}$ single crystal with unprecedented accuracy. It is found that the system shows the nearly isotropic $H_{c2}(0)$ ~ 48 T and the strong bending effect in $H_{c2}^{ab}(T)$. Detailed temperature-dependence of the $H_{c2}$ curves could be successfully explained by the Werthamer-Helfand-Hohenberg (WHH) prediction that considers both the Pauli limiting and the spin-orbit scattering effects together. Our results suggest that the Pauli limiting effect can be a main source of the peculiar isotropic $H_{c2}(0)$ for this iron-chalcogenide superconductor.

## II. EXPERIMENTS

Single crystals of FeTe$_{0.6}$Se$_{0.4}$ were grown by the self-flux method in an evacuated quartz tube. The mixture of Fe and (Te,Se) with starting composition Fe(Te$_{0.6}$Se$_{0.4}$) was heated at 1193 K for 12 hours and slowly cooled down afterwards with the rate of 40 K / hour to room temperature. The resistivity was measured by standard four-probe method in a physical property measurement system (PPMS) up to 14 T and in a hybrid magnet (NHMFL, Tallahassee, USA) from 11.5 to 45 T down to 1.5 K. Two pieces of crystal with a rectangular shape were prepared by cleaving along the *ab*-plane in a single crystal piece. Both pieces, which confirmed to show the same $T_c$, were loaded onto a sample platform for the PPMS or the hybrid magnet for measuring their resistivity under high magnetic fields applied parallel to the *ab*-plane and *c*-axis, respectively.

## III. RESULTS AND DISCUSSION

Figure 1 shows the temperature-dependence of resistivity and magnetization. It is noted that resistivity of our sample shows the metallic behavior approximately below 200 K and shows superconductivity at 14.5 K, when determined from the 50 % of normal state resistivity. Its small transition width ($T_{onset} - T_{\rho=0}$) about 1.3 K supports the high quality of the crystal investigated. According to a recent study by Liu et al.,[19] $Fe_{1.12}Te_{0.72}Se_{0.33}$ with a large amount of excess Fe shows a semiconducting behavior down to $T_c$ while $Fe_{1.04}Te_{0.72}Se_{0.28}$ with less Fe shows the metallic temperature-dependence. The amount of interstitial Fe existing between the FeTe(Se) layer was suggested to be a decisive factor to determine this contrasting transport behavior. Therefore, the metallic resistivity in our sample suggests that our sample is close to the stoichiometric $Fe(Te_{0.6}Se_{0.4})$ with minimal amount of excess, interstitial Fe. The diamagnetic signal of magnetic susceptibility $\chi$ measured at $H = 10$ Oe (inset of Figure 1) indicates the existence of bulk superconductivity with its volume fraction about 80 %.

Temperature-dependence of resistivity under static magnetic fields are summarized in Figure 2 for (a) $H // ab$ and (b) $H // c$. The temperature at which the zero-resistivity is realized is systematically suppressed with increasing magnetic field. The broadness of transition was pronounced in the resistivity curves for $H // c$, which is likely to result from the enhanced thermally activated vortex motion along this direction.[8] Moreover, at $H < 14$ T, the transition into the superconducting state for $H // ab$ occurs at higher temperatures than for $H // c$, indicating $H_{c2}^{ab}$ is higher than $H_{c2}^{c}$ at temperatures near $T_c$. On the other hand, at $H = 45$ T, the zero resistivity state is realized at 3.5 K for $H // c$, which is slightly higher than the corresponding value for $H // ab$, i.e., 2.8 K, indicating $H_{c2}^{c}$ is very close to or even slightly higher than $H_{c2}^{ab}$ near zero temperature. The magnetic field-dependence of resistivity is also plotted from 38 to 45 T at several fixed temperatures in Figure 3. Consistent with behavior seen in the temperature-dependence, the transition width becomes broader for $H // c$. Moreover, at $T = 12.3$

K, the superconducting state is obviously more stable for $H \parallel ab$ than for $H \parallel c$, while at $T = 3.7$ K, transition into the normal state occurs almost at the same $H$.

We determined the temperature-dependent $H_{c2}^{ab}$ and $H_{c2}^{c}$ curves from the resistivity data summarized in Figure 2 and 3. To determine superconducting transition temperature or fields from the resistivity, we choose the criterion that 50 % of the normal state resistivity is realized at $T_c$. With this criterion, we could minimize the effects of the vortex motion expected from the 10 % criterion or superconducting fluctuation expected from the 90 % criterion. As shown in Figure 4, the resultant $H_{c2}^{ab}$ and $H_{c2}^{c}$ curves from the both field-and temperature-sweeps well overlap each other, showing consistency between the two experimental methods to determine the $H_{c2}$ curves.

The $H_{c2}$ curves show anisotropic behavior near $T_c$, but become progressively isotropic as temperature is lowered; $\gamma \equiv H_{c2}^{ab}/H_{c2}^{c}$ is about 3 near $T_c$ and 0.99 at $T = 3.8$ K. Thus, it is likely that $H_{c2}(0)$ is nearly isotropic and approximately reaches ~ 48 T. Therefore, our results clearly show that nearly isotropic $H_{c2}(0)$ is realized even in our iron-chalcogenide superconductor, suggesting in turn it is a common physical feature in both '122' and '11' systems.[7, 9, 18] In the former, the isotropic $H_{c2}(0)$ was observed in both hole- and electron-doped single crystals as well as in a thin film, indicating the isotropic behavior is less sensitive to doping level and degree of disorder. Combining the previous case of observing nearly isotropic $H_{c2}(0)$ in an Fe excessive crystal of $Fe_{1.11}Te_{0.6}Se_{0.4}$ (Ref. 19) and our present results in a more stoichiometric $FeTe_{0.6}Se_{0.4}$, it is inferred that the isotropic $H_{c2}$ property is also robust against the variation of excess Fe doping level. This observation strongly suggests that the isotropic $H_{c2}(0)$ property might not be a simple consequence of the three dimensional band nature coming from the band warping effect in the apparently cylindrical Fermi-surfaces.

A noteworthy feature seen in the $H_{c2}^{ab}(T)$ curves is the existence of quite steep increase of $H_{c2}$ near $T_c$ and subsequent flattening of the curvature at lower temperatures. The calculated maximum slope -$dH_{c2}^{ab}/dT_c |_{max}$ ~ 13 T/K corresponds to the largest among the reported values in the iron-based superconductors. This is a key feature noticed also by Kida et al.[17] in the $H_{c2}(T)$ curve of a polycrystalline $FeTe_{0.75}Se_{0.25}$ sample. On the other hand, in a recent $H_{c2}^{ab}(T)$ study of a single crystal $Fe_{1.11}Te_{0.6}Se_{0.4}$ specimen by Fang et al.[18], the flattening feature in the $H_{c2}^{ab}(T)$ has not been clearly identified possibly due to the lack of data points near $T_c$. From the steeply increasing slope of the $H_{c2}^{ab}$ and $H_{c2}^{c}$ curves in Figure 4, we can calculate the orbital limiting fields for each crystallographic direction. According to the WHH formula predicting the orbital limiting field $H_{c2}^{orb}$ for a BCS superconductor with a single active band,[20] $H_{c2}^{orb}(0) = -0.69\ dH_{c2}/dT|_{T=T_c} T_c$, thus yielding $H_{c2}^{orb}(0) = $ 131.6 T in $H // ab$ and 56.5 T in $H // c$. These calculated values of $H_{c2}^{orb}(0)$ are much larger than the observed $H_{c2}(0) \sim 48$ T, suggesting that the low temperature $H_{c2}$ is predominantly Pauli-limited upper critical field. The expected Pauli limiting field for a weakly coupled BCS superconductor[21-22] is estimated as $H_P(0) \equiv 1.86\ T_c = 27.0$ T, which is much smaller than the predicted $H_{c2}^{orb}(0)$ as well as the experimental $H_{c2}(0) =\sim 48$ T. This observation implies that the spin-paramagnetic effect shall play an important role to determine $H_{c2}(0)$ in this '11' system and that a mechanism to enhance the Pauli limiting field beyond the BCS scenario might be also necessary. On the other hand, recent scanning tunneling microscopy studies in stoichiometric Fe(Te,Se) crystals report that the gap energy $\Delta$ closely matches with the BCS prediction of $2\Delta/k_B T_c \sim 3.5 - 3.8$. Thus, a simple scenario of the system being in a strongly coupled non-BCS regime with a larger gap than that expected from the mean field theory may not be adequate to explain the enhanced Pauli limiting field.[23-24]

Previous several reports on the $H_{c2}(T)$ in Fe-based superconductors showed that a two-band model in combination with orbital limiting effects can effectively describe the overall curvature of $H_{c2}$.[2-3, 7, 25-26] The main motivation of invoking the two band model is to explain the almost linear or sublinear increase with concave shape in the $H_{c2}^{c}$ curve near $T_c$ and its change to a convex form with decreasing temperature. However, in our case, both $H_{c2}^{ab}$ and $H_{c2}^{c}$ curves exhibit always the convex shape and the $H_{c2}^{ab}$ curve shows a flattening at temperatures below around $T_c/2$, all of which are not compatible with the expected $H_{c2}$ shape in the two-band system. Therefore, to describe the $H_{c2}$ curves of the present Fe-chalcogenide, the spin-paramagnetic effect should be taken into account in addition to the orbital pair-breaking effect, but not necessarily the multi-band effect. It is expected that the Pauli limiting will be quite effective in explaining the isotropic $H_{c2}(0)$ limit while the orbital limiting can explain the anisotropy between $H_{c2}^{ab}$ and $H_{c2}^{c}$ curves existing near $T_c$.

With this motivation, we attempted to fit the experimental $H_{c2}$ curves by the WHH formula that considers the spin-paramagnetic effect with the Maki parameter $\alpha$ in a single band system. Moreover, we also included the spin-orbit scattering constant $\lambda_{so}$ in the fitting.[20] For $H \parallel ab$, the attempt to fit the data without Pauli paramagnetic effect, i.e. $\alpha = 0$ only explained the experimental $H_{c2}(T)$ curve near $T_c$ and showed clear deviation at low temperatures (dotted line in Figure 4). The best fit was obtained when $\alpha = 5.5$ and $\lambda_{so} = 1.0$. The large value of $\alpha = 5.5$ is comparable to that of CeCoIn$_5$ and organic superconductors that have shown the first-order transition in $H_{c2}$ to form the Fulde–Ferrell–Larkin–Ovchinnikov (FFLO)-like state.[27-28] It is well known that such a large value of $\alpha = 5.5$ without a finite spin-orbit scattering $\lambda_{so}$ would cause a first order transition at low temperatures within the WHH formula, often interpreted as a possible realization of the FFLO-like states, contrary to our experimental curve. We chose the value of $\lambda_{so} = 1.0$ to avoid this transition and also to produce the best fit for the experimental $H_{c2}^{ab}$ over the broad temperature region, as demonstrated in Figure 4. The result of fitting indicates that a proper value of $\lambda_{so}$ is essential in determining the shape of the $H_{c2}^{ab}$ curves effectively. Moreover, the presence of a finite $\lambda_{so}$ made the predicted $H_{c2}^{ab}(0)$ enhanced over the $H_{c2}^{ab}(0)$ value with $\lambda_{so} = 0$ and the same $\alpha$, being consistent with the fact that the strong spin-orbit scattering suppresses the Pauli limiting effect.

On the other hand, for the experimental $H_{c2}^{c}$ curve, a relatively small $\alpha = 1.0$ (dotted line in Figure 4) was better in describing the $H_{c2}^{c}$ curve at low temperatures than the $\alpha = 0$ case (solid line in Figure 4), indicating the Pauli limiting effect exists for both directions. The fit results to the $H_{c2}^{c}$ data were not very sensitive to the variation of $\lambda_{so}$ from 0 to 3. Because there is no *a priori* reason that the orbital current would experience different spin-orbit scatterings for each crystallographic direction, we chose the same value of $\lambda_{so} = 1.0$ both for $H_{c2}^{c}$ and $H_{c2}^{ab}$. The theoretically predicted $H_{c2}(T)$ curves show good agreements with experimental data at overall temperatures except a small temperature-window between 8 and 11 K.

Summing up these fitting results, the presence of $\alpha$, the Make parameter, describing the Pauli limiting effect in the WHH scheme was essential to describe much smaller $H_{c2}(0)$ values than the expected orbital limiting field. Therefore, the Pauli limiting is postulated to be a dominant mechanism to determine the nearly isotropic $H_{c2}(0)$ behavior because the Zeeman splitting energy should be active to break the singlet Cooper pair in an isotropic manner regardless of detail electronic structure. In this picture, a small difference between $H_{c2}(0)$ values for both directions may be due to the presence of small difference in the Landé $g$-factor, rendering the Zeeman splitting energy for the two directions become slightly different. From the ratio of $H_{c2}^{ab}(0)/H_{c2}^{c}(0) = 0.96$, as extrapolated from Figure 4, $g^{ab}/g^{c}$ is thought to be about 0.96, which prompts experimental tests in future. Furthermore, the results of fitting particularly for $H_{c2}^{ab}$ in Figure 4 strongly indicate that the existence of spin-orbit scattering in the Fe(Te,Se) system can be a decisive physical process to enhance the $H_{c2}^{ab}(0)$ beyond the Pauli limiting field for a weak BCS superconductor ($H_P(0) \equiv 1.86\ T_c$) as well as to determine the temperature-dependent evolution of $H_{c2}$ curves at low temperatures.

Having established the importance of the Pauli limiting effect in the Fe-chalcogenide superconductors, we suspect that the non-stoichiometry effect such as excess Fe or Te(Se) vacancies can be one of important sources to make the orbital limiting field enhanced over the Pauli limiting field. The orbital limiting field, estimated by the value of $-dH_{c2}/dT$ near $T_c$, is inversely proportional to the Fermi velocity and the mean free path. According to a recent band calculation for FeSe, the Se vacancy tends to result in significantly enhanced density of states at the Fermi energy $N(E_F)$ and thus the effective mass.[29] Moreover, the presence of defects is likely to reduce the mean free path of the system. Both of these effects will be favorable for resulting in the enhanced $-dH_{c2}/dT$ near $T_c$ and as a result, the enhanced orbital limiting fields larger than the Pauli limiting field in the Fe(Te,Se) system. We also note that the $N(E_F)$ of Fe(Te,Se) system is found to be relatively large compared with those of other Fe-pnictides, according to recent first-principle calculations.[15, 30] Therefore, we suggest that combination of these two main mechanisms or one of them can be responsible for the manifestation of the Pauli paramagnetic effect in the Fe-chalcogenide superconductors.

Our observation suggests that similar effects may be also equally important for understanding nearly isotropic $H_{c2}(0)$ behaviors observed in many '122' systems with various dopants and doping levels although the multiband effect is still needed to properly explain the sublinear increase of $H_{c2}$ curves particularly near $T_c$. In this sense, for more complete description of the $H_{c2}$ curves in various Fe-pnictides or Fe-chalcogenides, it may be necessary to consider a more complete theoretical scheme

that considers both the multiband orbital and the Pauli paramagnetic effects simultaneously.[26]

## IV. CONCLUSIONS

In summary, we have determined detailed temperature dependence of upper critical fields in a FeTe$_{0.6}$Se$_{0.4}$ single crystal by use of a static magnetic field up to 45 T applied along the *ab*-plane and the *c*-axis. The Pauli paramagnetic effect is clearly evidenced by the clear flattening in the $H_{c2}$ curves along the *ab*-plane and also indicated by nearly isotropic $H_{c2}$ (0) ≈ 48 T for both directions. The enhanced effective mass and carrier scattering coming from the excess Fe or Te(Se) vacancies are argued to be responsible for the manifesting Pauli limiting effect.

## ACKNOWLEDEMENTS

This work is supported by Korean government through the NRL (M10600000238), GPP (K20702020014-07E0200-01410), and Basic Science Research Programs (2009-0083512). KHK and SHK were supported by LG Yeonam Foundation and Seoul R&BD (10543), respectively. Work at NHMFL was performed under the auspices of the NSF, the State of Florida, and the U. S. DOE.

**Figure captions**

FIG. 1. (a) (Color online) Temperature-dependence of resistivity of FeTe$_{0.6}$Se$_{0.4}$ single crystal under zero magnetic field in a broad temperature window. $T_c$ is estimated to be 14.5 K from the 50 % of the normal state resistivity. (inset) The change of DC magnetic susceptibility $\chi$ multiplied by $4\pi$ near $T_c$ relative to the value at 15 K, which was measured at $H$ = 10 Oe applied along the $ab$-plane after cooling (solid) and zero field cooling (dotted).

FIG. 2. (Color online) Temperature-dependence of resistivity under 0 to 14 T with a step of 1 T, and 38, 39, 41, 42, 43, 45 T along (a) $H$ // $ab$-plane and (b) $H$ // $c$-axis.

FIG. 3. (Color online) Magnetic field-dependence of resistivity at 3.7, 4.0, 4.5, 4.9, 5.9, 6.7, 7.9, 8.7, 9.7, 11.4 and 12.3 K along (a) $H$ // $ab$-plane and (b) $H$ // $c$-axis.

FIG. 4. (Color online) $H_{c2}(T)$ curves for $H$ // $ab$-plane and $H$ // $c$-axis (symbols) determined with the criterion of the 50 % of normal state resistivity. The filled (open) symbols are taken from the temperature (magnetic field)-dependence of resistivity. Dotted lines represent the WHH prediction with only orbital limiting effects considered ($\alpha = \lambda_{so} = 0$). Solid lines indicate the best fits to the WHH curve with the parameters $\alpha = 5.5$ and $\lambda_{so} = 1.0$ for $H$ // $ab$-plane and $\alpha = 1.0$ and $\lambda_{so} = 1.0$ for $H$ // $c$-axis.

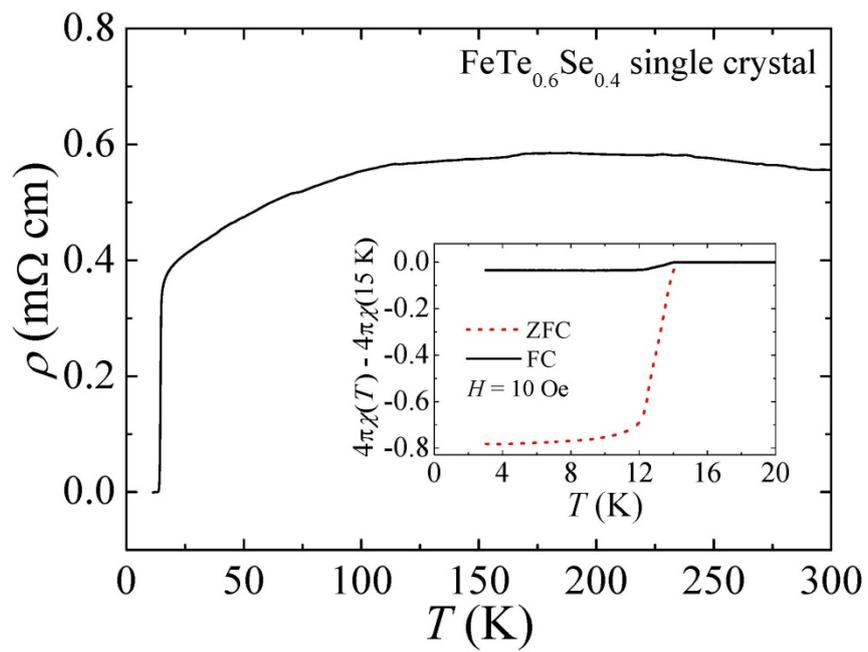

Fig 1. Seunghyun Khim *et al*.

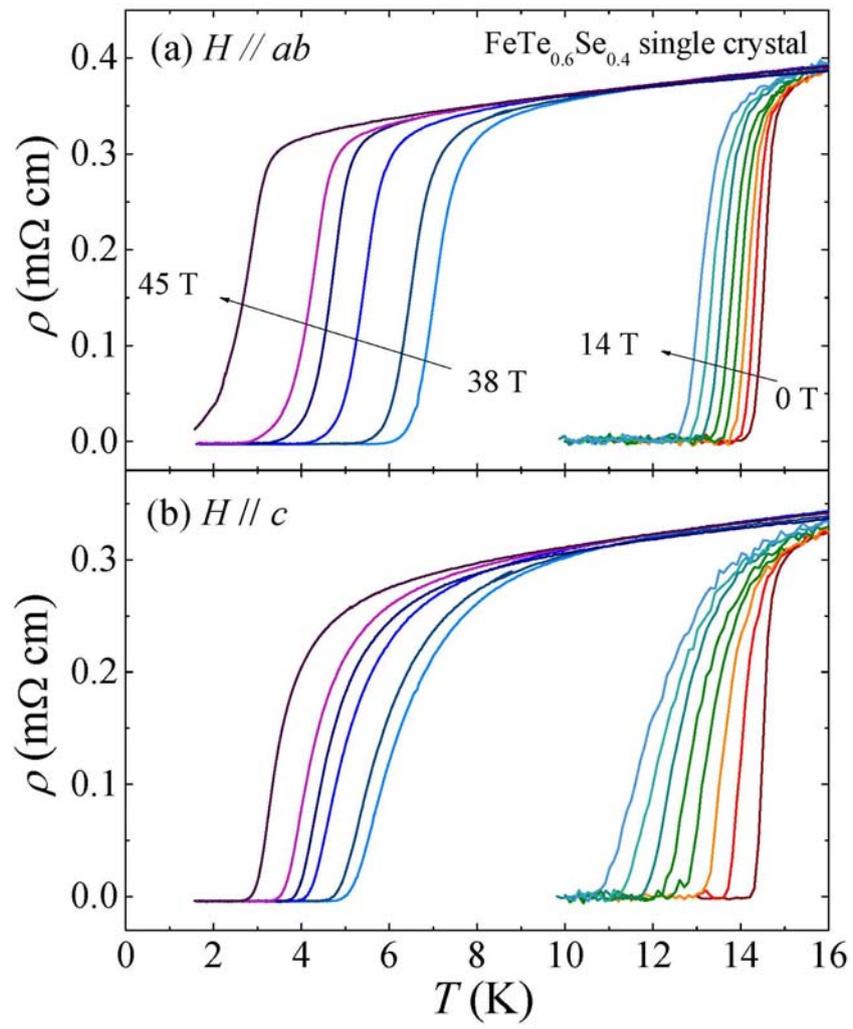

Fig 2. Seunghyun Khim *et al*.

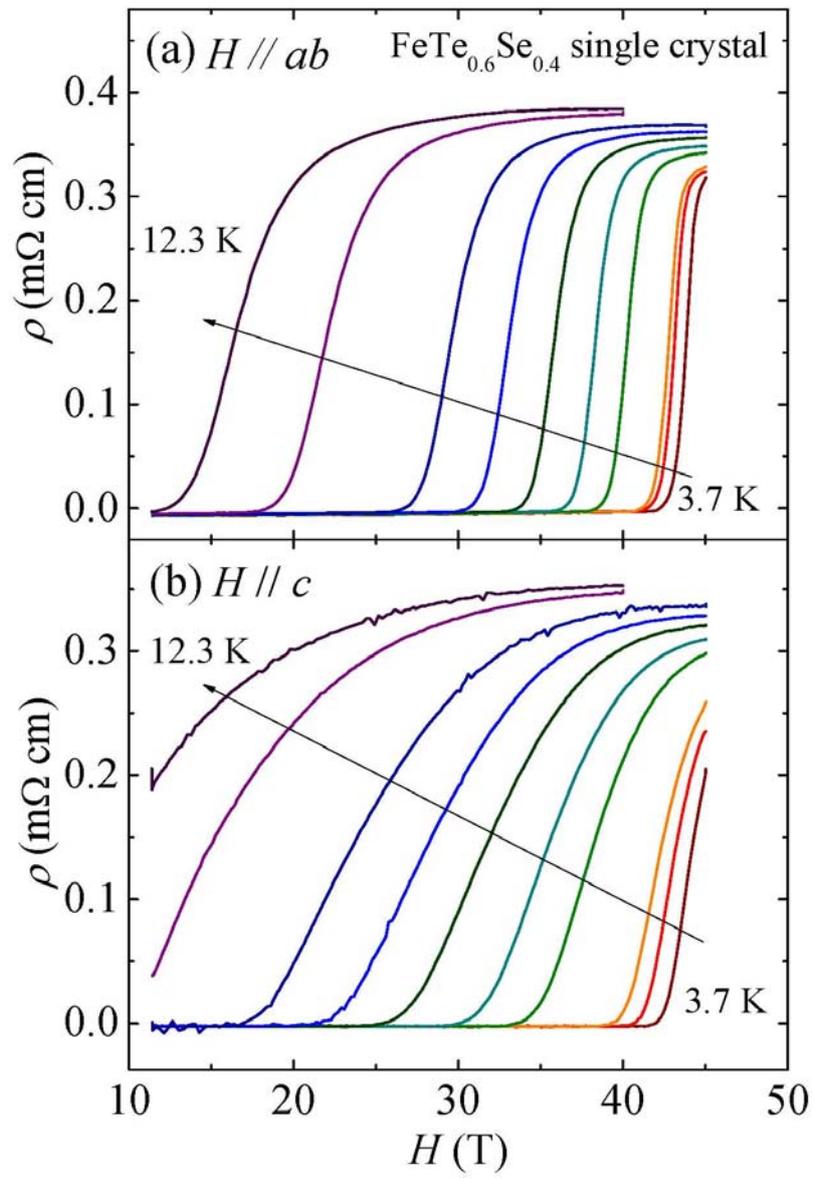

Fig 3. Seunghyun Khim *et al*.

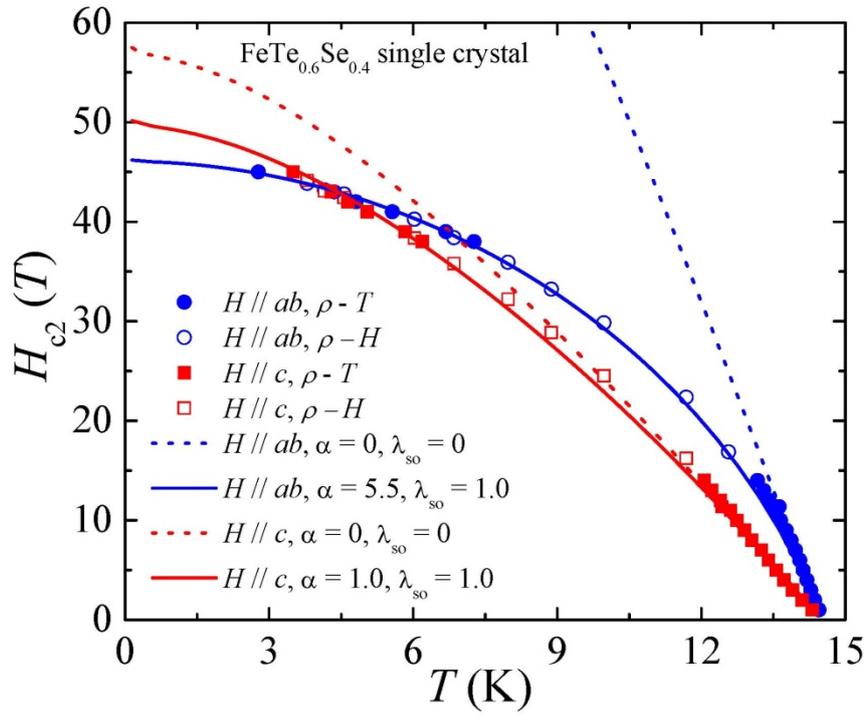

Fig 4. Seunghyun Khim *et al*.